\documentclass[traditabstract]{aa}
\usepackage[utf8]{inputenc}

\usepackage{graphicx}
\usepackage{rotating} 
\usepackage{lscape}
\usepackage{amssymb}
\usepackage{longtable}
\usepackage{xfrac}

\usepackage{natbib}
\bibpunct{(}{)}{;}{a}{}{,} 

\usepackage{gensymb} 

\begin{document}

\title{A temperate exo-Earth around a quiet M dwarf at 3.4 parsecs\thanks{Based on observations made with the HARPS instrument on the ESO 3.6 m telescope under the programme IDs 072.C-0488(A), 183.C-0437(A), and 191.C-0873(A) at Cerro La Silla (Chile). Radial velocity data (Table 5) are available in electronic form at the CDS via anonymous ftp to cdsarc.u-strasbg.fr (130.79.128.5) or via http://cdsweb.u-strasbg.fr/cgi-bin/qcat?J/A+A/}}
\author{X.~Bonfils   \inst{\ref{ipag}} 
   \and  N.~Astudillo-Defru \inst{\ref{obsge}}
   \and R.~D\'iaz \inst{\ref{uba},\ref{iafe}}
   \and J.-M.~Almenara \inst{\ref{obsge}}
   \and T.~Forveille \inst{\ref{ipag}}
   \and F.~Bouchy \inst{\ref{obsge}}
   \and X.~Delfosse \inst{\ref{ipag}}
   \and C.~Lovis \inst{\ref{obsge}}
   \and M.~Mayor \inst{\ref{obsge}}
   \and F.~Murgas \inst{\ref{iac},\ref{ull}}
   \and F.~Pepe \inst{\ref{obsge}}
   \and N.~C.~Santos \inst{\ref{uporto},\ref{caup}}
   \and D.~S\'egransan \inst{\ref{obsge}}
   \and S.~Udry \inst{\ref{obsge}}
   \and A.~W\"unsche \inst{\ref{ipag}}
}
        
\institute{Univ. Grenoble Alpes, CNRS, IPAG, 38000 Grenoble, France\label{ipag}
  \and Observatoire de Gen\`eve, Universit\'e de Gen\`eve, 51 ch. des Maillettes, 1290 Sauverny, Switzerland\label{obsge}
  \and Universidad de Buenos Aires, Facultad de Ciencias Exactas y Naturales. Buenos Aires, Argentina.\label{uba}
  \and CONICET - Universidad de Buenos Aires. Instituto de Astronom\'ia y F\'isica del Espacio (IAFE). Buenos Aires, Argentina.\label{iafe}
  \and Instituto de Astrof\'sica de Canarias (IAC), E-38200 La Laguna, Tenerife, Spain\label{iac}
  \and Dept. Astrof\'isica, Universidad de La Laguna (ULL), E-38206 La Laguna, Tenerife, Spain\label{ull}
  \and Instituto de Astrof\'isica e Ci\^encias do Espa\c{c}o, Universidade do Porto, CAUP, Rua das Estrelas, 4150-762 Porto, Portugal\label{uporto}
  \and Departamento de F\'isica e Astronomia, Faculdade de Ci\^encias, Universidade do Porto, Rua do Campo Alegre, 4169-007 Porto, Portugal\label{caup}
}
\date{Received 20 September 2017 / Accepted 26 October 2017}

\abstract{The combination of high-contrast imaging and high-dispersion spectroscopy, which has successfully been use to detect the atmosphere of a giant planet, is one of the most promising potential probes of the atmosphere of Earth-size worlds. The forthcoming generation of extremely large telescopes (ELTs) may obtain sufficient contrast with this technique to detect O$_2$ in the atmosphere of those worlds that orbit low-mass M dwarfs. This is strong motivation to carry out a census of planets around cool stars for which habitable zones can be resolved by ELTs, i.e. for M dwarfs within $\sim$5 parsecs. Our HARPS survey has been a major contributor to that sample of nearby planets. Here we report on our radial velocity observations of Ross 128 (Proxima Virginis, GJ447, HIP 57548), an M4 dwarf just 3.4 parsec away from our Sun. This source hosts an exo-Earth with a projected mass $m \sin i = 1.35 M_\oplus$ and an orbital period of 9.9 days. Ross 128 b receives $\sim$1.38 times as much flux as Earth from the Sun and its equilibrium ranges in temperature between 269~K for an Earth-like albedo and 213~K for a Venus-like albedo. Recent studies place it close to the inner edge of the conventional habitable zone. An 80-day long light curve from K2 campaign C01 demonstrates that Ross~128~b does not transit. Together with the All Sky Automated Survey (ASAS) photometry and spectroscopic activity indices, the K2 photometry shows that Ross~128 rotates slowly and has weak magnetic activity. In a habitability context, this makes survival of its atmosphere against erosion more likely. Ross~128~b is the second closest known exo-Earth, after Proxima Centauri~b (1.3 parsec), and the closest temperate planet known around a quiet star. The 15~mas planet-star angular separation at maximum elongation will be resolved by ELTs ($>$ 3$\lambda/D$) in the optical bands of O$_2$. }


\maketitle

\section{Introduction}

Clever observing strategies and techniques, together with technological progress, are moving comparative exoplanetology towards increasingly Earth-like planets. The coolest stars, in particular, offer clear observational advantages: compared to FGK stars, and everything else being equal, planets around M dwarfs have larger reflex motions, deeper transits (for well-aligned systems), and more favourable star-planet contrast ratios. This has long motivated our radial velocity (RV) search for planets around M dwarfs, which started with the discovery of the first planet around such a star \citep[GJ876b;][]{1998A&A...338L..67D}\footnote{also detected by \citet{Marcy:1998jv}}. This now amounts to almost 40 detections, which include a few Earth-mass planets and a few super-Earths located in the habitable zones of their host \citep[e.g.][]{2017A&A...602A..88A}. M dwarfs have also been the focus of several other planet searches with spectacular discoveries, including Proxima Cen b \citep{2016Natur.536..437A}, TRAPPIST-1 planets \citep{2017Natur.542..456G}, and LHS1140b \citep{2017Natur.544..333D}.

Considering their number and their well-characterised selection function, these detections provide us with statistical insights into planet formation \citep{2013A&A...549A.109B}. At the same time, many of these individual detections, and all the more so when the planetary properties such as liquid water might exist on their surface, call for follow-up studies to characterise their atmosphere and constrain their structure, composition, and chemistry.

For the subset of planets that transit, transmission and occultation spectroscopy are the characterisation methods of choice. James Webb Space Telescope (JWST) transmission spectroscopy of a few dozen coadded transits of the TRAPPIST-1 planets $b$, $c,$ and $d$, for instance, is expected to have sufficient sensitivity to detect O$_3$ in putative Earth-like atmospheres for these planets \citep{2016MNRAS.461L..92B}. This makes these planets strong candidates for a biomarker detection within the next few years, but one should remember that the TRAPPIST-1 star emits intense extreme ultraviolet (XUV) radiation and frequently produces powerful stellar flares, which together might have sterilised, if not completely stripped out, the atmospheres of at least its closer-in planets \citep{2017A&A...599L...3B, 2017ApJ...841..124V}. With stellar activity factored in, the quiet M dwarf LHS1140 and its temperate super-Earth become an appealing alternative. Both TRAPPIST-1 and LHS1140 have been given top priority for JWST Guaranteed Time observing.

Planets that do not transit are generally more difficult to characterise, but can be found closer to our Sun. This translates into both increased brightness and wider angular separation and the closest non-transiting exo-Earths might thus be amenable to characterisation. The maximum angular separation between Proxima Cen and its $b$ planet, for instance, is 37 milli-arcsec and can be resolved at visible wavelength by an 8 m class telescope. To match the daunting $10^{-7}$ planet-to-star contrast ratio, \citet{2017A&A...599A..16L} have proposed to couple the SPHERE extreme adaptive optics system and the ESPRESSO high-resolution spectrograph, which combines the contrast enhancements that one can achieve with high-resolution spectroscopy and high-contrast imaging \citep{2014Natur.509...63S, 2015A&A...576A..59S}. Under slightly optimistic assumptions, \citet{2017A&A...599A..16L} have concluded that a few dozen observing nights at the VLT would detect O$_2$, H$_2$O, and possibly CH$_4$, which
like TRAPPIST-1 represents a historic opportunity to detect biomarkers in the near future. Like TRAPPIST-1, however, Proxima Cen flares strongly and often,
which likewise challenges the habitability of its planet \citep{2016ApJ...829L..31D}. 

In that context, we report the detection of a planet orbiting a 2\sfrac{1}{2} times more distant but much quieter M dwarf, Ross~128. The planet is only slightly more massive than our Earth, is temperate, and orbits a very nearby, slowly rotating, quiet M dwarf.  We discuss the properties of the star in Sect.~\ref{sect:star}, present the data in Sect.~\ref{sect:data}, and use archive photometry in Sect.~\ref{sect:rot} to determine the stellar rotation period. In Sect.~\ref{sect:rv}, we analyse the RVs and demonstrate the presence of a planet and an additional periodicity likely caused by stellar activity. The final model parameters are derived from a Markov Chain Monte Carlo (MCMC)\ algorithm with Gaussian processes in Sect.~\ref{sect:model}. In Sect.~\ref{sect:concl}, we conclude that, although K2 photometry excludes transit, the low stellar activity and moderate distance from Earth make Ross 128 $b$ a good target for biomarker searches with forthcoming telescopes.

\section{\label{sect:star}Star}

Ross~128 entered the literature as the 128th entry in the \citet{1926AJ.....36..124R} catalogue of high-proper motion stars, and has since acquired denominations including Proxima Virginis, FY~Virginis, GJ~447, HIP~57548, and LHS~315. The spectral type of this object is M4 and, owing to its proximity, it is one of the brightest representatives of this subclass (V$^{\rm mag}$=11.15, J$^{\rm mag}$=6.51, H$^{\rm mag}$=5.95, K$^{\rm mag}$=5.65). With a distance of just 3.4 parsec ($\pi=295.80\pm0.54$~mas; GAIA \citeyear{Collaboration:2016gd}), Ross 128 is the closest star in the Virgo constellation ($\alpha$=11$^{\rm h}$47$^{\rm m}$44.4$^{\rm s}$, $\delta$=$+$00$^{\rm o}$48${\rm '}$16.4${\rm ''}$; Epoch=2000). Including brown dwarfs, it is the 13$^{th}$ closest (sub-)stellar system to the Sun. Ross 128 is moving towards us and will actually become our closest neighbour in just 71,000 years from now \citep[$D_{ca}=1.9$~pc;][]{2001A&A...379..634G}. 

 \citet{2015ApJ...804...64M} have derived its effective temperature $T_{\rm eff}=3192\pm60$~K, mass $M_\star=0.168\pm0.017$, radius $R_\star=0.1967\pm0.0077,$ and metallicity $[Fe/H]=-0.02\pm0.08$. Accordingly, its luminosity is $L_\star=0.00362\pm0.00039~L_\odot$. In \citet{2017A&A...600A..13A}, we measured a low \ion{Ca}{ii} emission level $\log (R'_{HK})=-5.573\pm0.082$. The calibration between $\log (R'_{HK})$ and the stellar rotation period $P_{rot}$ in the same paper converts this low calcium-line emission to an estimated rotation period of approximately 100 days, which is indicative of an age of the order of a few Gyr \citep{Newton:2016ea}. 

\begin{table}
\caption{
\label{table:stellar}
Observed and inferred stellar parameters for Ross 128. 
}
\begin{tabular}{l@{}lcc}
\hline
Spectral type$^{(1)}$        &                  & M4                    \\
Epoch$^{(2)}$              & & 2000\\
Right ascension, $\alpha^{(2)}$ && 11$^{\rm h}$47$^{\rm m}$44.3974$^{\rm s}$\\
Declination, $\delta^{(2)}$ && $+$00$^{\rm o}$48${\rm '}$16.395${\rm ''}$\\
Parallaxe, $\pi$$^{(2)}$       & [mas]        & $295.80\pm0.54$ \\
Distance, d$^{(2)}$                &[pc]           & $3.3806 \pm 0.0064  $    \\

\multicolumn{3}{l}{Stellar photometry}\\
~~~~V$^{(3)}$                            &[mag]         & $11.15$ \\
~~~~J$^{(4)}$                             & [mag]        & $6.505\pm0.023$ \\
~~~~H$^{(4)}$                           & [mag]        & $5.945\pm0.024$ \\
~~~~K$^{(4)}$                           &  [mag]        & $5.654 \pm 0.024$     \\
Effective temperature, T$_{\rm eff}^{(5)}$                 & [K]    & 3192$\pm$60\\
Mass, $M_\star^{(5)}$       & [M$_\odot$]             & 0.168$\pm$0.017\\
Radius, $R_\star^{(5)}$       & [R$_\odot$]             & 0.1967$\pm$0.0077\\
Metallicity, [M/H]$^{(5)}$                 &     & $-0.02\pm0.08$\\
Luminosity, $L_\star$       & [$\mathrm{L_\odot}$]    &  $0.00362\pm0.00039$\\
$\log(R'_{HK})^{(6)}$  && $-5.573\pm0.082$\\
Rotation period, $P_{Rot}^{(6,7,8)}$&  [days] & 101, 121, 123\\
Age, $\tau^{(9)}$ & [Gyr] & $\gtrsim$5 & \\
\hline
\end{tabular}\\
(1): \citet{Henry:2002cf}; (2): GAIA \citeyear{Collaboration:2016gd}; (3) \citet{Landolt:1992hl}; (4): \citet{Cutri:2003tp}; (5): \citet{2015ApJ...804...64M}; (6) \citet{2017A&A...600A..13A}; (7): this work using ASAS photometry (see Sect.~\ref{sect:rot}); (8): this work from RV (see Sect.~\ref{sect:rv}) (9) \citet{Newton:2016ea} given $P_{Rot}$; 
\end{table}

\section{\label{sect:data}Data}
From July 26, 2005 (BJD=2453578.46) to April 26, 2016 (BJD=2457504.7), we collected 157 observations with the HARPS spectrograph \citep{2003Msngr.114...20M, 2004A&A...423..385P}. Exposure times were fixed to 900 sec. We discarded the 158th measurement that appears in the ESO archives, which is a just a 5 second exposure (March 23, 2015; BJD=2456740.68).  We used the high-resolution mode (R=115'000), with the scientific fibre illuminated by the target and calibration fibre either unused or illuminated by the sky. The data reduction followed the same steps as in all our recent papers. Spectral extraction and calibration relied on the on-line pipeline \citep{2007A&A...468.1115L}, which also gives an initial guess for the RV. An offline processing then refines the RV measurements and their uncertainties \citep[e.g.][]{2015A&A...575A.119A, 2017A&A...602A..88A}. The line spread function changed significantly when the May 2015 upgrade of HARPS replaced its fibre link with octogonal fibres. In this work, we treat the pre- and post-upgrade data as independent time series, which appear in the figures  in red and blue, respectively. Table ~\ref{tab:rv} (only available electronically) gives the RV time series in the barycentric reference frame. Before proceeding to the next section, however, we removed the small but significant secular acceleration ($dRV/dt = 0.14 ~m/s/yr$), which we computed using the distance and proper motion of Ross 128 \citep[$\mu_\alpha=0.60526''/yr$, $\mu_\delta=-1.21926''/yr$;][]{vanLeeuwen:2007dc} and Eq. 2 of \citet{Zechmeister:2009kw}.


To complement our HARPS observations, we used archive photometry from both ASAS and K2. The All Sky Automated Survey \citep[ASAS;][]{1997AcA....47..467P} observed Ross 128 for over nine~years. We retrieved its V-band photometry extracted through the smallest ASAS aperture, ASAS MAG 0. The K2 mission \citep{Howell:2014ju} observed Ross 128 for 82 days in its Campaign 1. We retrieved  the K2 light curves detrended with the EVEREST \citep{Luger:2016fj} and POLAR \citep{Barros:2016hj} pipelines from the Mikulski Archive for Space Telescopes (MAST)\footnote{https://archive.stsci.edu/k2/}.

\begin{figure}
\includegraphics[width=\linewidth]{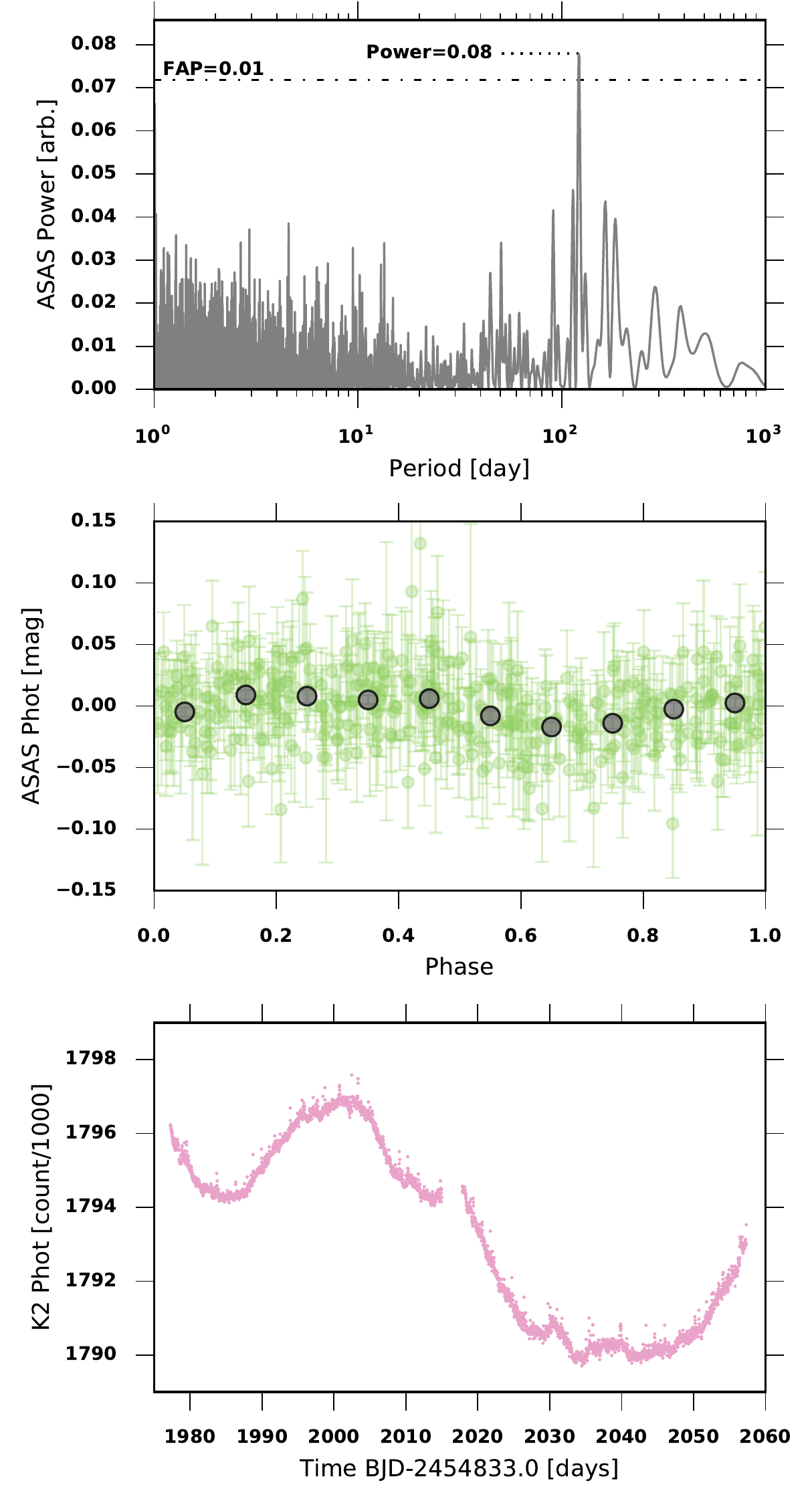}\\
\caption{\label{fig2a}Photometry of Ross 128. {\it Top:}  periodogram of the ASAS V-band photometry. {\it Middle :} ASAS photometry phase-folded to P=121.2 days. Grey filled circles are median values in 0.1-phase bins. {\it Bottom :} K2 photometry extracted with the Everest pipeline \citep{Luger:2016fj} as a function of time. We only show corrected photometry FCOR with the highest quality flag.}
\end{figure}

\section{\label{sect:rot}Stellar rotation}
Since inhomogeneities such as spots, plages, or inhibition of the convection at the surface of a rotating star can induce apparent Doppler shifts, prior knowledge of the stellar rotation helps eliminate false positive planets. The low $\log (R'_{HK})$ of Ross 128 already indicates that its rotation period is long, ${\sim}100$ days. Here, we used ASAS and K2 photometry to refine its value. 

We only retained the last seven years of the more than nine years of ASAS photometry, since Ross 128 was sampled infrequently prior to BJD$=$2452500. We subtracted the median value of each observing season, clipped out all $4\sigma$ outliers, and computed the generalised Lomb-Scargle periodogram \citep[GLS; ][]{2009A&A...496..577Z}. As seen in Fig.~\ref{fig2a} (top panel), the GLS has obvious power excess for periods around 121 days with power $p_{\rm max}=0.08$. We evaluated the power threshold for a given false alarm probability (FAP) on virtual data sets generated by bootstrap with replacement. The 1\% FAP threshold is $p_{1\%}=0.07$ and the 121~days periodic signal is therefore significant. The phased photometry (middle panel of Fig.~\ref{fig2a}) shows a $\sim1\%$ semi-amplitude.

The K2 photometry has orders of magnitude better precision than the ground-based measurements and provides quasi-continuous observations during 80 days, but does not cover a full stellar rotation. The $\sim$0.4\% trend of this photometry over 80~days is compatible with ${\sim}1\%$ variations on a 121~day period (bottom panel of Fig.~\ref{fig2a}).

\section{\label{sect:rv}Evidence for an orbiting planet and additional stellar activity}
\begin{figure}
\includegraphics[width=\linewidth]{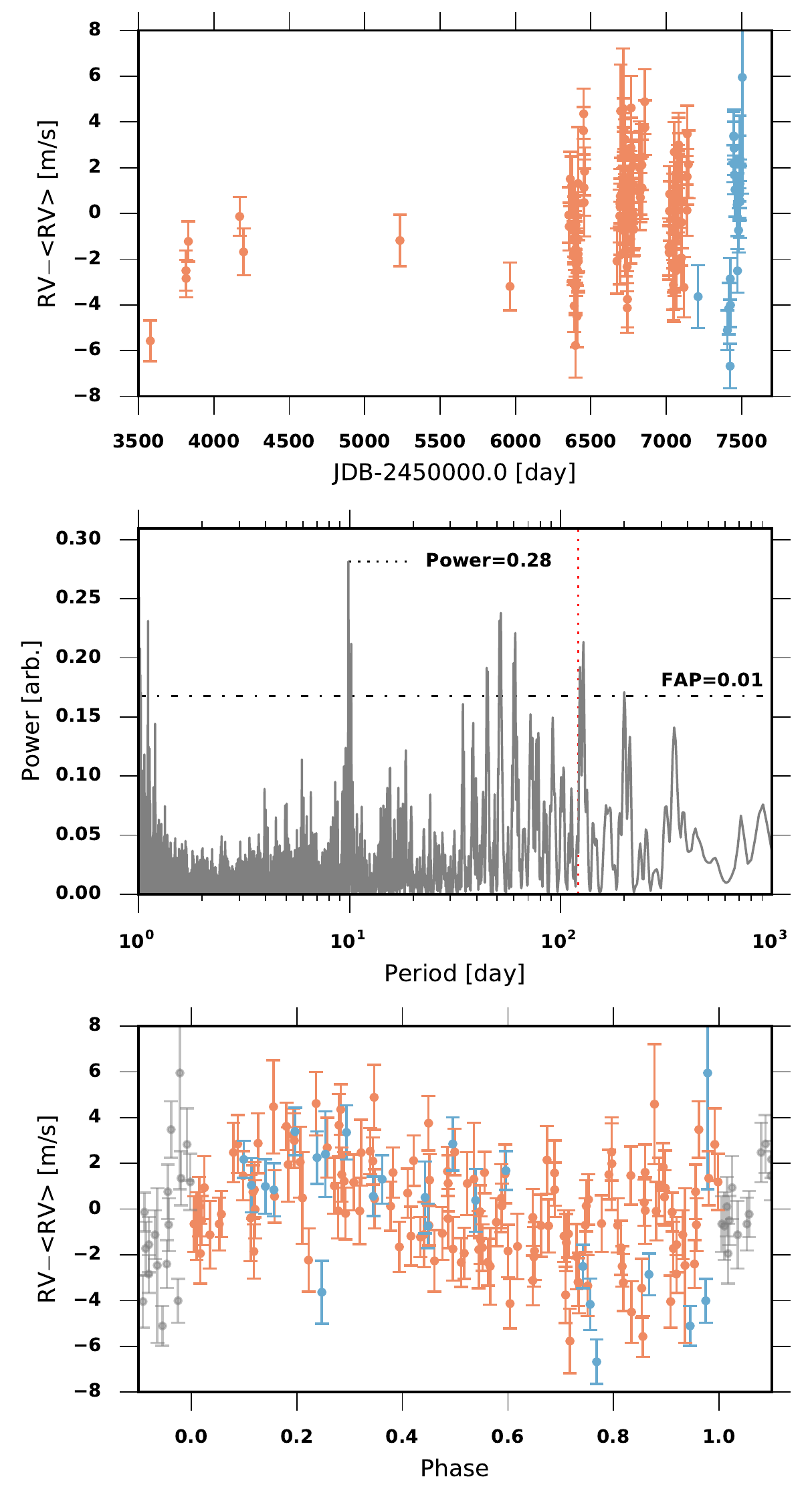}\\
\caption{\label{fig1a}HARPS radial velocities: as a function of time ({\it top}), periodogram ({\it middle}) and phase-folded to P=9.9 days ({\it bottom}). The red and blue points represent pre- and post-upgrade measurements, respectively. }
\end{figure}

The raw pre- and post-upgrade RV time series (Fig.~\ref{fig1a}, top panel) have r.m.s. dispersions of 2.1 and 3.0~m/s, respectively, i.e. well in excess of the $\sim1.2$ m/s expected from the photon noise on the individual measurements, and a constant model has a Bayesian information criterion BIC=618. The GLS periodogram shows a prominent power excess around period of 9.9 days and several other significant peaks (Fig.~\ref{fig1a}, middle panel). The maximum power $p_{\rm max}=0.28$ is well in excess of the 1\% FAP threshold $p_{1\%}=0.17$ and the detection of a periodic signal is thus highly significant. For a visual sanity check, we phase the RVs to a 9.9 day period (Fig 1, bottom panel) and see that the signal is well sampled at every phase. 

The 9.9~day period is comfortably away from the 121~day stellar rotation period (Sect.~\ref{sect:rot}) and its first few harmonics, which by itself already lends considerable confidence to its interpretation as a planet detection. A Levenberg-Marqardt adjustment of a Keplerian model has r.m.s. residuals of 1.9 and 2.6 m/s for the two time series and an overall BIC=429. The planet's orbital period is $P_b=9.9$~ days,  RV semi-amplitude is $K_1=1.7$~m/s, and eccentricity is compatible with zero. 

\begin{figure}
\includegraphics[width=\linewidth]{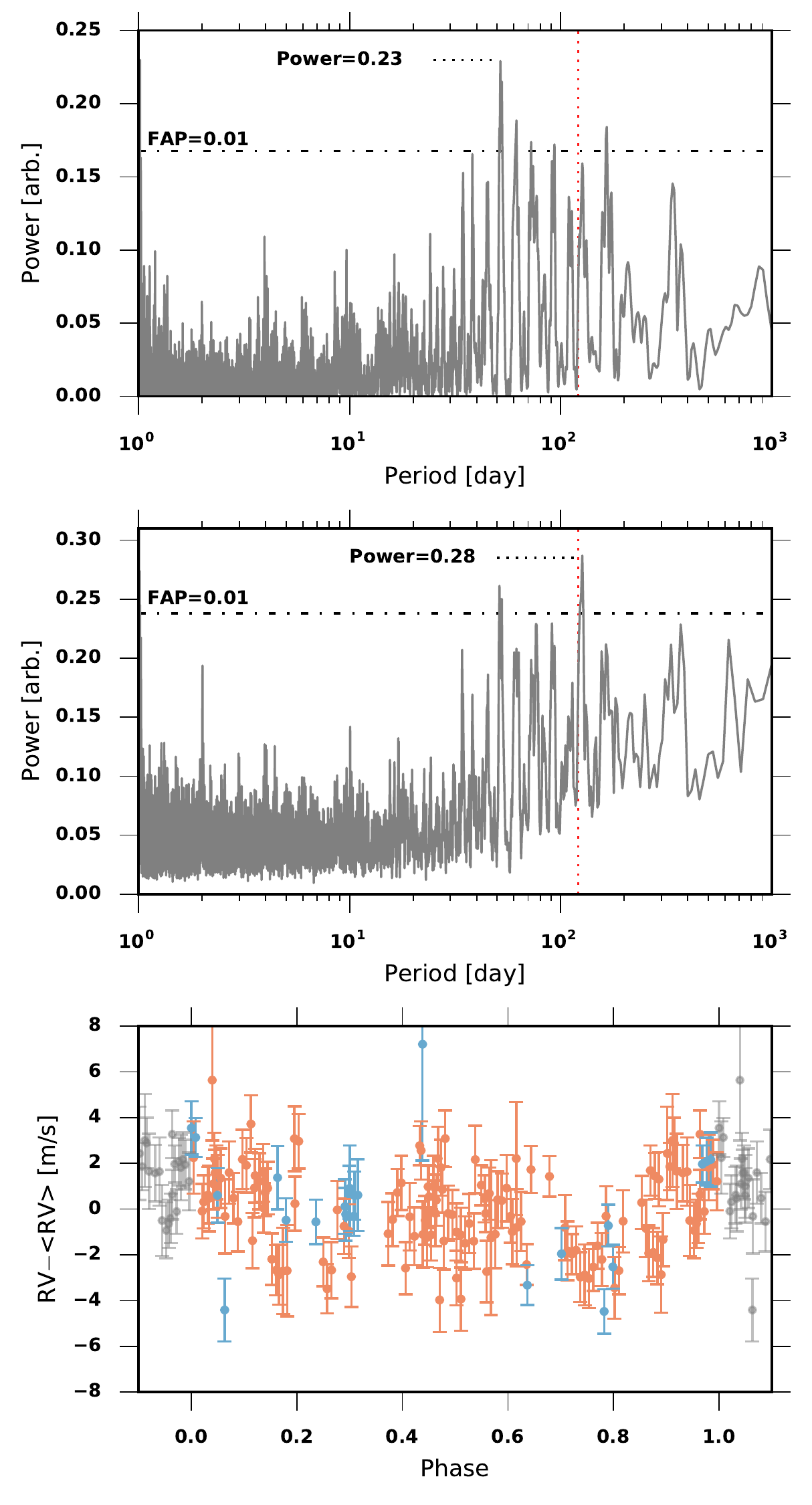}\\
\caption{\label{fig3a}Radial velocity residuals after subtraction of the best-fit Keplerian and drift. {\it Top:} GLS periodogram is shown. {\it Middle}: KGLS periodogram is shown. {\it Bottom:} Residuals phase-folded to a $P=123$~days period are shown. The vertical red dashed line indicates the 121~day rotation period inferred from the ASAS photometry. The red and blue points represent pre- and post-upgrade measurements, respectively.}
\end{figure}

The residuals from the 9.9~day Keplerian model thus remain well in excess of the dispersion expected from pure photon noise, and we searched for periodicities in those residuals using both GLS and Keplerian-GLS \citep[KGLS; ][]{2009A&A...496..577Z} periodograms.  Whereas the GLS measures the power of a sine fit at each period, the KGLS does so for Keplerian signals, therefore exploring periodicities for a wider range of functional shapes. The GLS (Fig.~\ref{fig3a}, top panel) has its maximum power, $p_{max}=0.23$, at a $P=51.8$~day period, and multiple other peaks above the 1\% FAP threshold. The $\sim$52 days peak remains significant in the KGLS (Fig.~\ref{fig3a}, middle panel), but the most powerful peak ($p_{max}=0.28$) is now at $P=123$~days. This period is very close to the $\sim$~121 day stellar rotation period inferred from the ASAS photometry (Sect. 4), and is certainly compatible with it after accounting for the effect of differential rotation. The residuals phase-folded with a $P=123$~day period (bottom panel of Fig.~\ref{fig3a}) suggest a coherent signal at that period with an approximate symmetry around phase 0.5. This approximate symmetry predicts excess power in the 123/2~day second harmonic of the rotation period, as is indeed observed, and the $\sim$52~day peak of the GLS additionally matches a 1~year alias of 123/2 days. The power excess in the residuals of the Keplerian fit is therefore entirely consistent with two-spotted stellar activity modulated by stellar rotation with a $\sim$120 day period. Since the spot configuration is likely to have evolved over the $\sim$11~years of the HARPS measurements, the next section models the effect of this configuration using Gaussian process regression rather than a deterministic physical model. 

\begin{figure}
\includegraphics[width=\linewidth]{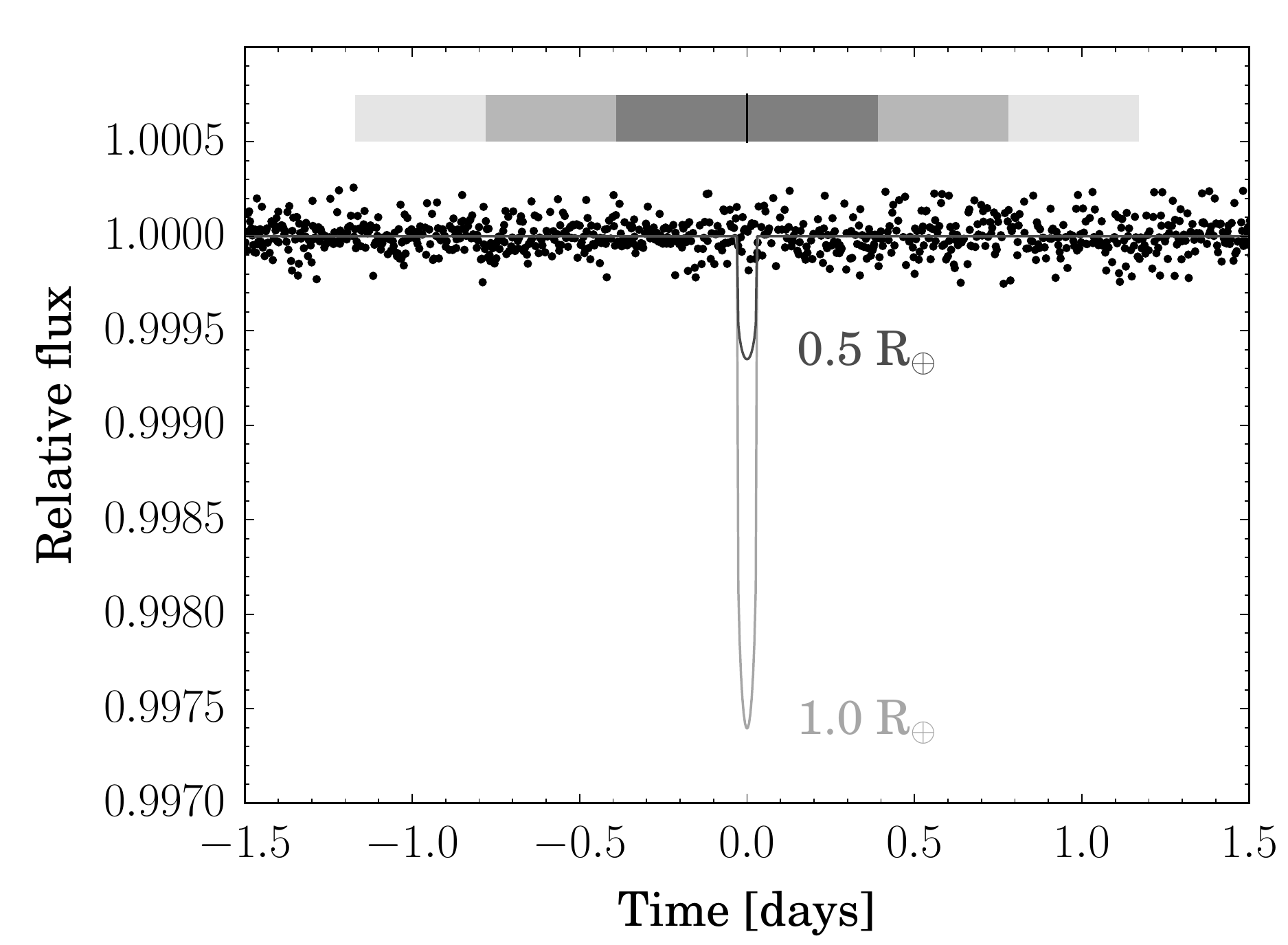}\\
\caption{\label{fig4a}Transit search in the K2 photometry.}
\end{figure}

\section{\label{sect:model}Modelling}

\newcommand{\struct}{\epsilon}
\newcommand{\per}{\mathcal{P}}

Our model of the HARPS RVs consists of a single Keplerian function representing the effect of the planetary companion, and we explore the effect of including an additional distant body modelled as a linear velocity drift. Since the RV time series contains additional signals with frequencies close to the rotational rate of the star, its harmonics, and aliases, we modelled the error term as a multivariate Gaussian distribution with a covariance matrix produced by an appropriate kernel function. This includes the effect of correlation between the data points into the model.

For the kernel function, we chose a quasi-periodic kernel, 
$$
k_\mathrm{QP}(t_i, t_j) = A^2 \exp\left(-\frac{(t_i - t_j)^2}{2\tau^2} - \frac{2}{\epsilon}\sin^2{\left(\frac{\pi (t_i - t_j)}{\mathcal{P}}\right)}\right)\;,
$$
which is known to represent adequately the covariance produced by active regions rotating in and out of view \citep[e.g.][]{Haywood:2014hs, 2015MNRAS.452.2269R}. This kernel function has four hyperparameters, corresponding to the amplitude of the covariance term ($A$),  rotational period of the star ($\mathcal{P}$), covariance decay time ($\tau$), and shape parameter ($\epsilon$). To test the robustness of our results with respect to the choice of kernel function, we also explored models employing the simpler squared-exponential kernel,
$$
k_\mathrm{SE}(t_i, t_j) = A^2 \exp\left(-\frac{1}{2}\frac{(t_i - t_j)^2}{\tau^2}\right)\;,
$$
with only two hyperparameters, $A$ and $\tau$, and no periodic term. In addition, an extra white noise component was added to the model by adding the following term to each of the previous kernels:
$$
k_\mathrm{WN}(t_i, t_j) = \delta_{ij} \left[\sigma^2_i + S_i \sigma^2_J + S^+_i \left(\sigma^+_J\right)^2\right]\;,
$$
where $\delta_{ij}$ is the Kronecker delta function, $\sigma_i$ is the internal incertitude of the data point taken at time $t_i$; $\sigma_J$ and $\sigma^+_J$ are the width of the additional noise component for the pre-, and post-upgrade data, respectively; and $S_i$ is an indicator variable, whose value is one if observation $i$ is taken before the HARPS fibre upgrade and zero otherwise, and vice versa for $S^+_i$.

In summary, four models were tested and were constructed by combining the two variants for the data model, i.e. a single Keplerian (k1) or a Keplerian plus a linear velocity drift (k1d1), and the two options for the noise term model, i.e the squared-exponential kernel (sek) and quasi-periodic kernel (qpk). For the Bayesian inference of the model parameters, we set the priors listed in Table~\ref{table.priors}. The pre- and post-upgrade velocities were treated independently with a different extra white noise amplitude for each, and an offset between these velocities. 

The model parameters were sampled using the MCMC algorithm described in \citet{Goodman+10} and implemented by \citet{2013ascl.soft03002F}. The initial positions of 300 walkers were randomly drawn from the prior distribution. The algorithm was run for 40'000 iterations, and the walkers were evolved to different posterior maxima. The separate maxima were identified by clustering the samples in parameter space and the marginal likelihood of each mode was estimated using the importance sampling estimator described by \citet{2013arXiv1311.0674P}. In all cases, a single mode exhibited overwhelming evidence with respect to all other secondary maxima. The walkers in the secondary maxima were then replaced by new walkers initiated in the main maximum and the algorithm was run until no further evolution of the samplers was seen. This step took between 15'000 and 40'000 iterations, depending on the model. Then 100'000 additional steps were run, on which the final inference was performed. 

\begin{figure*}
\center
\includegraphics[width=0.9\columnwidth]{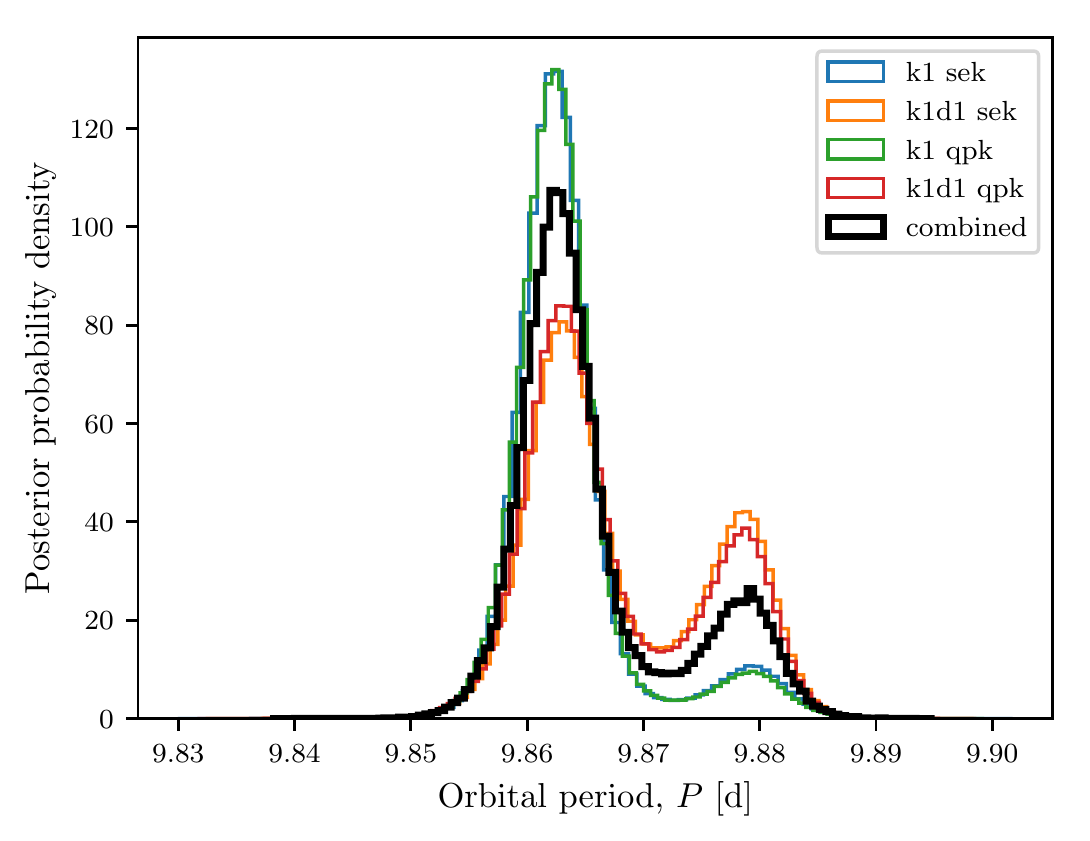}
\includegraphics[width=0.9\columnwidth]{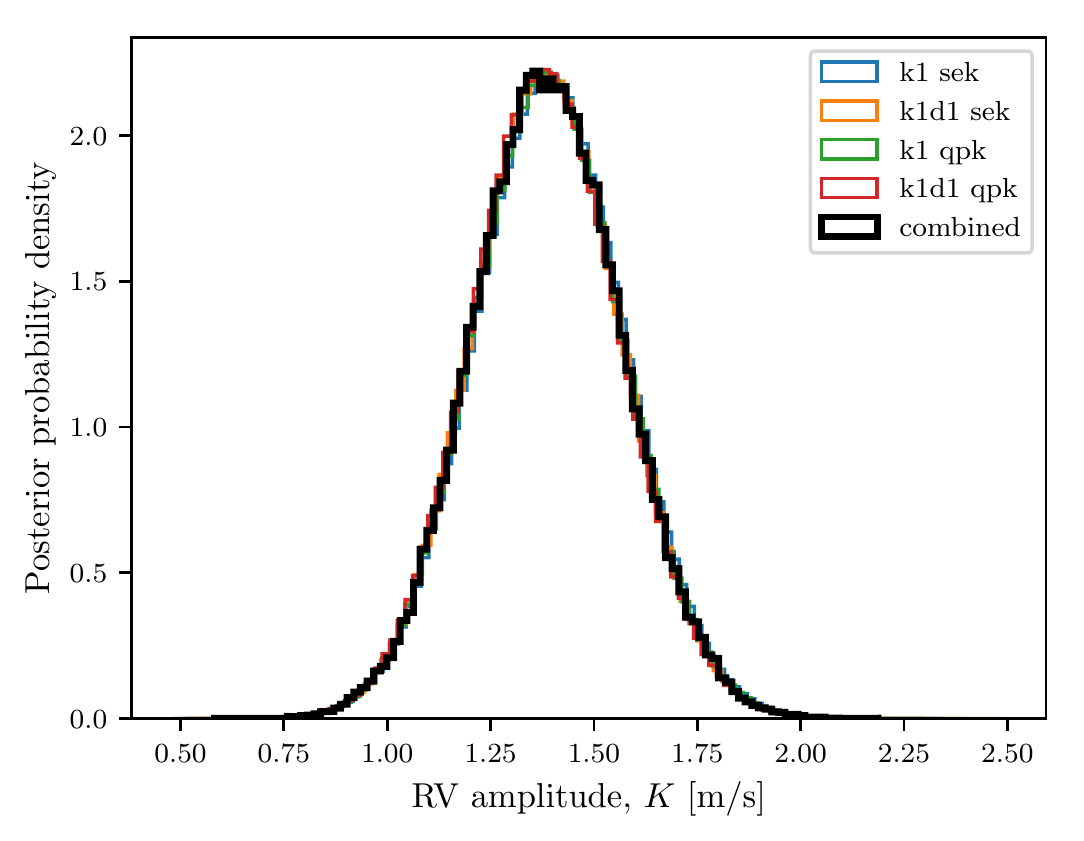}
\caption{Marginal posterior distribution of the orbital period (\emph{left}) and RV semi-amplitude (\emph{right}) for the four tested models and the weighted average of all four. The period marginal posterior distribution has a more marked bimodality for models including a non-zero acceleration term.}
\label{fig.posteriors}
\end{figure*}

Results are reported in Table~\ref{table.resultsallmodels} for each tested model. The inferred results on most planet parameters are independent of the choice of model (see e.g. the marginal posterior of the velocity amplitude in Fig.~\ref{fig.posteriors}). The most notable exception is the orbital period, which exhibits a bimodal distribution, with modes centred on 9.86 and 9.88 days, in which the difference relative weight of the modes depends on whether the model includes a linear drift or not (Fig.~\ref{fig.posteriors}). The other parameters that change slightly with the inclusion of a linear drift are the velocity zero-point, the mean longitude at epoch, the amplitude of the covariance, $A$, and the offset between pre- and post-upgrade velocities. The evolution timescale hyperparameter $\tau$ marginal distribution varies significantly between models with different kernel functions.

\begin{figure}
\center
\includegraphics[width=0.9\columnwidth]{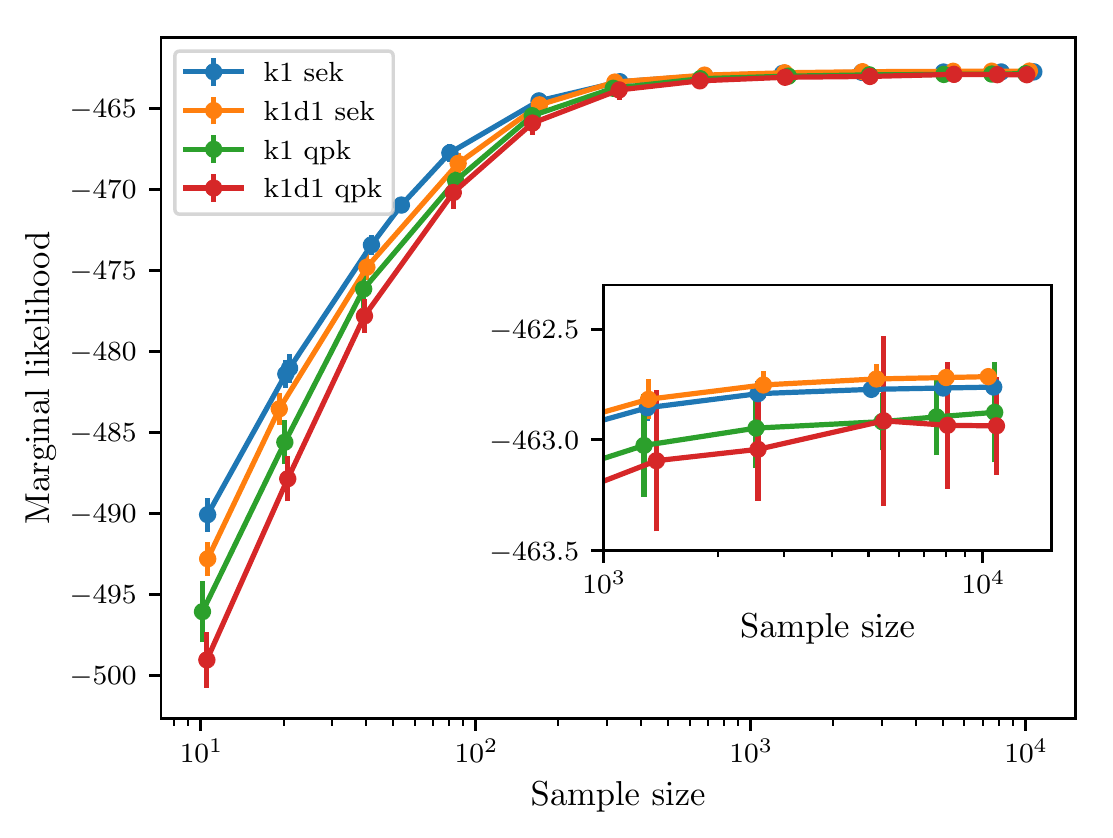}
\caption{Evolution of the marginal likelihood estimation for the four tested models with sample size. The known bias of the importance sampling estimator used \citep{2013arXiv1311.0674P} is evident, but seems to become negligible for sample sizes larger than around 2000. The inset shows an enlargement of the region with sample sizes between 1000 and 10'000, where a slight preference for the squared exponential models is seen. A small random noise was added in the x-direction to facilitate viewing.}
\label{fig.evidences}
\end{figure}

\begin{figure}
\center
\includegraphics[width=0.9\columnwidth]{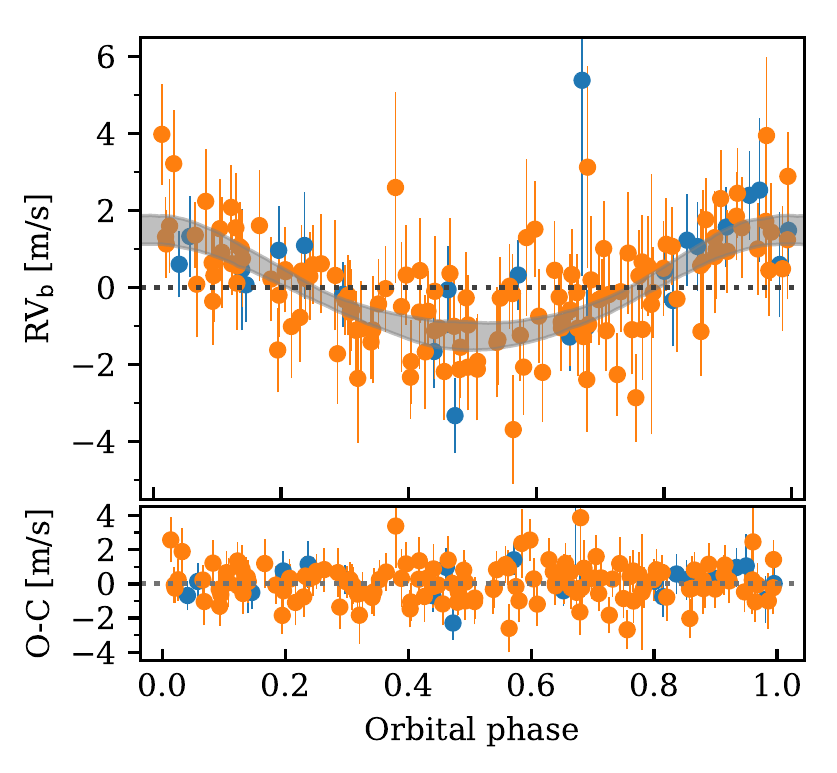}
\caption{Radial velocity induced by the planet on its host star, phase-folded to the maximum-a posteriori (MAP) orbital period. The effect of the secular acceleration and the Gaussian process prediction were subtracted from the data (using the MAP parameter values). The orange and blue points correspond to the pre- and post-upgrade velocities, respectively. The inferred MAP additional white noise term ($\sigma_J$ and $\sigma_J^+$) were added in quadrature to the velocity uncertainties. The grey shaded region extends between the 5th and 95th percentile of the model at each orbital phase, computed over 10,000 samples of the merged posterior distribution.}
\label{fig.phasefolded}
\end{figure}

The relative merits of each model was studied by estimating the marginal likelihood of each model using the importance sampling estimator of \citet{2013arXiv1311.0674P}. This is a biased estimator, so we explored the evolution of the estimation for each model as the size of the sample increased (Fig.~\ref{fig.evidences}). After around 5000 samples, the estimator seems to have converged. All models are approximately equally good at explaining the data with a slight preference for the squared-exponential kernel. The final inference on the model parameters was carried out by combining the samples from the four models weighted by their posterior probability; this probability was, in turn, computed assuming all four tested models form an exhaustive set, i.e. their probabilities add up to the value one. The results are listed in table~\ref{table.results} and the MAP velocity model is presented in Fig.~\ref{fig.phasefolded}.

\newcommand{\jeff}[2]{\mathcal{J}(#1, #2)}
\newcommand{\mjeff}[2]{\mathcal{MJ}(#1, #2)}
\newcommand{\unif}[2]{\mathcal{U}(#1, #2)}

\begin{table}[h]
\caption{\label{table.priors}Prior distribution for the model parameters.
  $\unif{x_{min}}{x_{max}}$ is the uniform distribution and $\jeff{x_{min}}{x_{max}}$ is the Jeffreys distribution (log-flat) between $x_{min}$ and $x_{max}$. $\mathcal{N}(\mu, \sigma)$ is the normal distribution with mean $\mu$ and scale $\sigma$, and $\mjeff{a_0}{x_{max}}$ is the modified Jeffreys distribution\tablefootmark{a}.}
    \centering
    \begin{tabular}{lc}
    \hline\hline\noalign{\smallskip}
         Parameter \& units & Prior distribution\\
         \noalign{\smallskip}
         \hline
         \noalign{\smallskip}
         \multicolumn{2}{c}{\bf Zero-point, offset and drift}\\
         \hline\noalign{\smallskip}
         $\gamma_0$\tablefootmark{b} [m/s] & $\unif{-20}{20}$\\
         
         $\gamma_1$\tablefootmark{c} [m/s/yr] & $\mathcal{N}(0, 3)$\\
         
         $\delta_{12}$ [m/s] &$\unif{-10}{10}$\\
         
         \noalign{\smallskip}
         \hline\noalign{\smallskip}
         \multicolumn{2}{c}{\bf Noise model parameters}\\
         \hline\noalign{\smallskip}

         $\sigma_J$ [m/s]& $\mjeff{1}{10}$\\

         $\sigma^+_J$  [m/s]& $\mjeff{1}{10}$\\

         A  [m/s]& $\mjeff{1}{10}$\\

         $\log\tau$ [days]& $\unif{1}{3}$\\

         $\per$\tablefootmark{d} [days]& $\jeff{1}{1000}$ \\

         $\epsilon$\tablefootmark{d} & $\unif{0.5}{10}$ \\

         \hline\noalign{\smallskip}
         \multicolumn{2}{c}{\bf Planet parameters}\\
         \hline\noalign{\smallskip}         

         $P$ [days]& $\jeff{1}{100}$ \\

         $K$ [m/s]& $\mjeff{1}{10}$ \\

         $\sqrt{e}.\sin({\omega})$ & $\unif{-1}{1}$\\

         $\sqrt{e}.\cos({\omega})$ & $\unif{-1}{1}$\\

         $\lambda_0$ & $\unif{-\pi}{\pi}$\\

         $e$ & $\unif{0}{1}$\\

         \hline\noalign{\smallskip}
    \end{tabular}
    \tablefoot{\\
      \tablefoottext{a}{The modified Jeffreys distribution is defined as $$f(a_0, x_{max}; x)\mathrm{d}x = \frac{\mathrm{d}x}{a_0\left(1 + x/a_0\right)}\frac{1}{\log\left(1 + x/a_0\right)}\;\;.$$}\\
      \tablefoottext{b}{Around HARPS mean velocity, $-30.8907$ km/s.}\\
      \tablefoottext{c}{Only in models with a linear drift.}\\
      \tablefoottext{d}{Only in models with quasi-periodic kernel.}
      }
    \label{table.priors}
\end{table}

\begin{landscape}

\begin{table}
\caption{\label{table.resultsallmodels}Orbital elements inferred from each tested model. Values correspond to the posterior sample mean, with errors being the standard deviation of the MCMC samples. In the second line we report the 95\% highest density interval (HDI). }


\begin{tabular}{llcccc}
\hline
\hline\noalign{\smallskip}
&& \multicolumn{4}{c}{\bf Models}\\ 
&& k1 sek & k1d1 sek & k1 qpk & k1d1 qpk\\ 
\hline\noalign{\smallskip}
\multicolumn{6}{l}{\bf Noise model parameters}\\ 
\hline\noalign{\smallskip}
$\sigma_J$ & [m/s] & $0.14 \pm 0.13$ & $0.13 \pm 0.13$ & $0.85 \pm 0.13$ & $0.04 \pm 0.13$  \\ 
&& $[0.00, 0.43]$& $[0.00, 0.43]$& $[0.00, 0.42]$& $[0.00, 0.43]$ \\ 
\noalign{\smallskip}
$\sigma^+_J$ & [m/s] & $1.07 \pm 0.35$ & $0.43 \pm 0.35$ & $0.15 \pm 0.35$ & $0.14 \pm 0.35$  \\ 
&& $[0.00, 1.17]$& $[0.00, 1.12]$& $[0.00, 1.15]$& $[0.00, 1.12]$ \\ 
\noalign{\smallskip}
A & [m/s] & $1.76 \pm 0.34$ & $1.78 \pm 0.31$ & $1.76 \pm 0.36$ & $1.90 \pm 0.33$  \\ 
&& $[1.38, 2.73]$& $[1.19, 2.46]$& $[1.33, 2.78]$& $[1.22, 2.53]$ \\ 
\noalign{\smallskip}
$\log_{10}\tau$ & [d] & $1.155 \pm 0.092$ & $1.144 \pm 0.097$ & $1.46 \pm 0.63$ & $1.57 \pm 0.62$  \\ 
&& $[0.957, 1.315]$& $[0.904, 1.295]$& $[1.03, 2.31]$ $\cup$ $[2.36, 3.00]$& $[0.98, 2.34]$ $\cup$ $[2.44, 3.00]$ \\ 
\noalign{\smallskip}
$\per$ & [d] & --& --& $110.0 \pm 223.2$ & $254.7 \pm 222.6$  \\ 
&& & & $[60.9, 840.2]$& $[41.0, 840.2]$ \\ 
\noalign{\smallskip}
$\struct$ &  & --& --& $1.4 \pm 2.8$ & $3.3 \pm 2.8$  \\ 
&& & & $[0.7, 1.5]$ $\cup$ $[1.8, 10.0]$& $[0.9, 1.6]$ $\cup$ $[1.8, 10.0]$ \\ 
\noalign{\smallskip}
\hline\noalign{\smallskip}
\multicolumn{6}{l}{\bf Planet parameters}\\ 
\hline\noalign{\smallskip}
Orbital period, $P$ & [d] & $9.8596 \pm 0.0056$ & $9.8634 \pm 0.0078$ & $9.8650 \pm 0.0054$ & $9.8824 \pm 0.0077$  \\ 
&& $[9.8544, 9.8702]$ $\cup$ $[9.8759, 9.8830]$& $[9.8562, 9.8825]$& $[9.8542, 9.8700]$ $\cup$ $[9.8761, 9.8822]$& $[9.8558, 9.8825]$ \\ 
\noalign{\smallskip}
RV amplitude, $K$ & [m/s] & $1.34 \pm 0.18$ & $1.51 \pm 0.18$ & $1.28 \pm 0.18$ & $1.31 \pm 0.18$  \\ 
&& $[1.05, 1.78]$& $[1.05, 1.77]$& $[1.01, 1.75]$& $[1.02, 1.76]$ \\ 
\noalign{\smallskip}
Orbital eccentricity, $e$ & [0..1[ & $0.036 \pm 0.092$ & $0.030 \pm 0.090$ & $0.065 \pm 0.090$ & $0.044 \pm 0.089$  \\ 
&& $[0.000, 0.299]$& $[0.000, 0.295]$& $[0.000, 0.291]$& $[0.000, 0.290]$ \\ 
\noalign{\smallskip}
Mean longitude, $\lambda_0$ & [deg] & $68.0 \pm 7.7$ & $74.9 \pm 8.5$ & $73.6 \pm 7.8$ & $76.2 \pm 8.6$  \\ 
&& $[62.9, 93.6]$& $[62.5, 96.9]$& $[61.8, 92.8]$& $[62.3, 97.3]$ \\ 
\noalign{\smallskip}
\noalign{\smallskip}
$\sqrt{e} \cos{\omega}$ &  & $-0.08 \pm 0.21$ & $0.10 \pm 0.21$ & $0.25 \pm 0.21$ & $0.11 \pm 0.21$  \\ 
&& $[-0.20, 0.58]$& $[-0.21, 0.56]$& $[-0.23, 0.55]$& $[-0.23, 0.55]$ \\ 
\noalign{\smallskip}
$\sqrt{e} \sin{\omega}$ &  & $0.17 \pm 0.19$ & $-0.14 \pm 0.19$ & $0.01 \pm 0.19$ & $0.18 \pm 0.19$  \\ 
&& $[-0.41, 0.33]$& $[-0.39, 0.34]$& $[-0.40, 0.33]$& $[-0.38, 0.35]$ \\ 
\noalign{\smallskip}
Minimum mass, $M_p \sin i$ & [$\mathrm{M}_\oplus$] & $1.35 \pm 0.20$ & $1.49 \pm 0.20$ & $1.31 \pm 0.20$ & $1.30 \pm 0.20$  \\ 
&& $[0.98, 1.80]$& $[0.98, 1.81]$& $[0.99, 1.83]$& $[1.00, 1.82]$ \\ 
\noalign{\smallskip}
Semi-major axis, $a$ & [AU] & $0.0493 \pm 0.0017$ & $0.0487 \pm 0.0017$ & $0.0497 \pm 0.0017$ & $0.0492 \pm 0.0017$  \\ 
&& $[0.0460, 0.0529]$& $[0.0462, 0.0532]$& $[0.0462, 0.0528]$& $[0.0459, 0.0529]$ \\ 
\noalign{\smallskip}
\hline\noalign{\smallskip}
\multicolumn{6}{l}{\bf Planet parameters}\\ 
\hline\noalign{\smallskip}
Systemic velocity, $\gamma_0$ & [km/s] & $-30.89022 \pm 0.00048$ & $-30.88978 \pm 0.00049$ & $-30.89032 \pm 0.00053$ & $-30.89056 \pm 0.00053$  \\ 
&& $[-30.89175, -30.88976]$& $[-30.89125, -30.88914]$& $[-30.89189, -30.88970]$& $[-30.89135, -30.88913]$ \\ 
\noalign{\smallskip}
Linear drift, $\gamma_1$ & [m/s/year] & --& $0.48 \pm 0.15$ & --& $0.47 \pm 0.15$  \\ 
&& & $[0.06, 0.68]$& & $[0.06, 0.68]$ \\ 
\noalign{\smallskip}
Velocity offset, $\delta_{12}$ & [m/s] & $-3.6 \pm 1.1$ & $-3.0 \pm 1.2$ & $-3.5 \pm 1.2$ & $-3.9 \pm 1.2$  \\ 
&& $[-4.1, 0.5]$& $[-5.3, -0.4]$& $[-4.2, 0.8]$& $[-5.6, -0.6]$ \\ 
\noalign{\smallskip}
\hline
\end{tabular}

\end{table}

\end{landscape}


\begin{table*}
\label{table.results}
\center
\caption{Orbital elements inferred from the model mixture. Values correspond to the posterior sample mean, with errors being the standard deviation of the MCMC samples. In the second line we report the 95\% highest density interval (HDI). Only parameters common to all models are tabulated.}
\label{table.results}
\begin{tabular}{llc}
\hline\noalign{\smallskip}
N$_{\rm Meas}$&& 159\\
\noalign{\smallskip}
\hline
\noalign{\smallskip}
$\gamma$ & [km/s] & $-30.89946 \pm 0.00058$\\
&&$[-30.89166, -30.88927]$\\
\noalign{\smallskip}
$\delta_{12}$ & [m/s]& $-2.4 \pm 1.3$\\
&&$[-5.1, 0.2]$\\
\noalign{\smallskip}\hline\noalign{\smallskip}
\multicolumn{3}{c}{Noise model parameters}\\
\noalign{\smallskip}\hline\noalign{\smallskip}
$\sigma_J$ & [m/s]   & $0.17\pm 0.13$ \\
                &&$[0.00, 0.43]$\\
\noalign{\smallskip}
$\sigma^+_J$ & [m/s] & $0.46\pm 0.35$\\
                &&$[0.00, 1.13]$\\
\noalign{\smallskip}
$A$ & [m/s] & $1.91\pm 0.35$ \\
                &&$[1.26, 2.62]$\\
\noalign{\smallskip}\hline\noalign{\smallskip}
\multicolumn{3}{c}{\bf Ross 128 b}\\
\noalign{\smallskip}\hline\noalign{\smallskip}
  $P$& [days]&    $9.8658\pm0.0070$ \\
                &&$[9.85582, 9.87111] \cup [9.87330, 9.88313]$\\
\noalign{\smallskip}
$K$& [m/s]&    $1.39\pm0.18$ \\
            && $[1.01, 1.74]$ \\
\noalign{\smallskip}
  $e$& [0..1[&    $0.116\pm0.097$ \\
            && $[0.000, 0.304]$ \\
\noalign{\smallskip}
$\lambda_{0}$& [deg]&    $78.2\pm8.4$ \\
BJD$_{\rm ref}$=2456740  &&  $[61.92, 95.32]$    \\
\noalign{\smallskip}\hline\noalign{\smallskip}
$\sqrt{e}.\cos({\omega})$&  &   $0.19\pm0.22$ \\
                            &&  $[-0.21, 0.56]$ \\
\noalign{\smallskip}
$\sqrt{e}.\sin({\omega})$&  &   $-0.04\pm0.20$ \\
                            &&  $[-0.41, 0.31]$ \\
\noalign{\smallskip}
M $\sin$ i& [M$_{\rm Earth}$]&   $1.40\pm0.21$ \\
                           && $[0.99, 1.83]$ \\
\noalign{\smallskip}
  a& [au]&                  $0.0496\pm0.0017$ \\
        &&      $[0.0461, 0.0528]$ \\
\noalign{\smallskip}
  T$_{eq}$ for A$_B$=[0.75, 0]& [K] & [213, 301] \\
\noalign{\smallskip}
  BJD$_{\rm Trans}-2456740$&[days]  & $0.07\pm 0.39$ \\
        &&      $[-0.79, 0.78]$ \\
\noalign{\smallskip}
\hline
\end{tabular}
\end{table*}

\section{\label{sect:concl}Discussion}

The $m \sin i = 1.35~m_\oplus$ Ross~128~b planet orbits Ross~128 with a 9.86~day period, and at 0.049~AU is $\sim20$ times closer to its star than the Earth is to the Sun. Since the star is $\sim280$ times less luminous than the Sun, Ross 128~$b$ receives just 1.38 times more energy than our Earth. For assumed albedos of 0.100, 0.367, or 0.750, its equilibrium temperature would thus be 294, 269, or 213 K. Using theoretically motivated albedos, the \citet{2017ApJ...845....5K} criteria place the planet firmly outside the habitable zone, while \citet{2013ApJ...765..131K}, \citet{2014ApJ...787L...2Y}, and \citet{2016ApJ...819...84K} find it outside, inside and just at the inner edge of the habitable zone. The precise location of the inner edge is therefore still uncertain, as it depends on subtle cloud-albedo feedbacks and on fine details in complex GCM models. The habitable zone most likely will not be firmly constrained until liquid water is detected (or inferred) at the surface of many planets. Meanwhile, it is probably preferable to refer to Ross 128~$b$ as a temperate planet rather than as a habitable zone planet. 

A planet just 3.4~parsecs away either having liquid water or just shy of having some makes an extremely appealing characterisation target. From the occurrence rate of temperate planets measured by Kepler, \citet{2015ApJ...807...45D} estimated that the closest habitable zone planet that transits its star is approximately 11~parsecs away. Yet, a stroke of luck could certainly align a closer temperate planet to undergo transits from our position in space, and all RV detections are therefore worth following up with photometry. As for Ross~128~$b$, existing K2 photometry readily answers whether it transits or not. We phase-folded the de-trended, low-frequency filtered, POLAR K2 photometry \citep{Barros:2016hj} to the ephemeris computed in the previous section (Fig.~\ref{fig4a}).  Ross~128~$b$ unfortunately does not transit, with central transits of any planet bigger than 0.19 R$_\oplus$ excluded at least at the 99\% confidence level. Non-grazing transits of a more realistic 0.5- or a 1.0-R$_\oplus$ planet are excluded with very high confidence.

Transit spectroscopy being excluded, we turn to the potential of measuring phase curves. \citet{2017AJ....154...77S} estimated that five days of JWST observations could detect the putative atmosphere of Proxima~Cen~b \citep[see also][]{2016ApJ...832L..12K}. Ross 128~b is not as favorable however, since its host star is 1.4 times larger and, at near- or mid-infrared wavelengths, $3-4$ times fainter than Proxima Centauri. Similar JWST observations for Ross 128~b are thus likely to be prohibitively expensive. 

The best odds of characterising Ross~128~b are most likely through combining the contrast improvements achieved with high-angular resolution and with high-spectral dispersion \citep{2002ApJ...578..543S}. \citet{2015A&A...576A..59S} investigated the potential of this strategy for rocky planets around our nearest neighbours and found that a putative temperate exo-Earth orbiting Proxima Cen could be detected in just 10 hours on the European ELT (E-ELT). A year later, \citep{2016Natur.536..437A} detected an actual planet with very similar properties using RV measurements, and the technique was immediately contemplated to characterise that planet. \citet{2016arXiv160903082L} proposed to upgrade the SPHERE adaptive optics system of the VLT and inject light from the location of the planet into the ESPRESSO high-resolution spectrograph to detect the planet in few tens of nights, and possibly detect its atmospheric O$_2$ with 60 nights of observations. Ross~128~b again is not quite as favorable as Prox~Cen~b, since it cannot be resolved by a 10~m-class telescope. Its 15~mas angular separation, however, will be resolved by the 39~m E-ELT at optical wavelengths ($>3\lambda/D$ in the O$_2$ bands) and its expected contrast is similar to that of Prox~Cen~b, owing to their similar radii and semi-major axes. The two host stars have similar optical apparent magnitudes, leading to similar planetary apparent magnitudes. A realistic investment of E-ELT resources can therefore most likely detect Ross~128~b with high-angular resolution plus high-dispersion spectroscopy, although not as easily as Prox~Cen~b.

On the flip side, Ross~128 is one of the quietest stars to host a temperate exo-Earth. \citet{2017ApJ...834...85N} measured an H$_\alpha$ equivalent width EW~H$_\alpha=-0.068~\AA$ which makes Ross~128 one the most quiescent M dwarfs. They classified stars as active when EW H$_\alpha < -1 \AA$ and, for comparison, measured EW H$_\alpha = -4.709 \AA$ for Proxima Cen b. Stellar activity is probably the highest concern regarding the emergence of life, and even the survival of an atmosphere, on planets orbiting M dwarfs. Restricting the target list to quiet stars would disqualify Proxima~Cen~b and leave Ross 128 b as the best temperate planet known to date. This will certainly make this new temperate exo-Earth a top target for characterisation with the ELTs.

\begin{acknowledgements}
We wish to thank the referee for his remarks that led to an improved manuscript. XB, JMA, and AW acknowledge funding from the European Research Council under the ERC Grant Agreement n. 337591-ExTrA. 
N.C.S. acknowledges the support from Funda\c{c}\~ao para a Ci\^encia e a Tecnologia (FCT) through national funds and by FEDER through COMPETE2020 by grants UID/FIS/04434/2013\&POCI-01-0145-FEDER-007672 and PTDC/FIS-AST/1526/2014\&POCI-01-0145-FEDER-016886. N.C.S. was also supported by FCT through Investigador FCT contract reference IF/00169/2012/CP0150/CT0002. 
\end{acknowledgements}

\bibliographystyle{aa}
\bibliography{mybib}
\onecolumn
\longtab{5}{\begin{longtable}{llll}
\caption{\label{tab:rv} Radial velocity time series of Ross 128, given in the barycentric reference frame of the solar system. The secular acceleration due to Ross 128's proper motion is not removed. Fourth column indicates whether data were collected before (1) or after (2) the HARPS fiber upgrade.}\\
\hline\hline
BJD-2400000.0 & RV [km/s] & $\sigma_{\rm RV}$[km/s]  & Instr. \\ \hline \endfirsthead
\caption{continued.}\\
\hline\hline
BJD-2400000.0 & RV [km/s] & $\sigma_{\rm RV}$[km/s]  & Instr. \\ \hline \endhead \hline \endfoot
53578.459205	&	-30.89613	&	0.00089	&	1\\
53814.759070	&	-30.89297	&	0.00088	&	1\\
53815.772088	&	-30.89331	&	0.00082	&	1\\
53830.713844	&	-30.89168	&	0.00088	&	1\\
54170.722180	&	-30.89046	&	0.00085	&	1\\
54196.763646	&	-30.89200	&	0.00102	&	1\\
55233.790844	&	-30.89109	&	0.00113	&	1\\
55963.836166	&	-30.89281	&	0.00105	&	1\\
56353.830518	&	-30.88954	&	0.00121	&	1\\
56356.786912	&	-30.89003	&	0.00105	&	1\\
56363.759284	&	-30.88795	&	0.00118	&	1\\
56373.666619	&	-30.88824	&	0.00127	&	1\\
56385.560103	&	-30.89119	&	0.00139	&	1\\
56386.590515	&	-30.89128	&	0.00115	&	1\\
56387.719406	&	-30.89053	&	0.00103	&	1\\
56388.639447	&	-30.89016	&	0.00119	&	1\\
56389.629716	&	-30.89349	&	0.00114	&	1\\
56390.632899	&	-30.89019	&	0.00148	&	1\\
56391.724678	&	-30.88946	&	0.00127	&	1\\
56393.680865	&	-30.88953	&	0.00145	&	1\\
56394.638315	&	-30.89063	&	0.00129	&	1\\
56395.633389	&	-30.89138	&	0.00112	&	1\\
56396.628948	&	-30.89108	&	0.00143	&	1\\
56397.608592	&	-30.89522	&	0.00141	&	1\\
56398.606879	&	-30.89266	&	0.00125	&	1\\
56399.601854	&	-30.89100	&	0.00121	&	1\\
56400.586042	&	-30.88995	&	0.00108	&	1\\
56401.563689	&	-30.89130	&	0.00119	&	1\\
56402.585338	&	-30.89168	&	0.00138	&	1\\
56408.621117	&	-30.89394	&	0.00135	&	1\\
56409.620951	&	-30.89190	&	0.00338	&	1\\
56410.604605	&	-30.89056	&	0.00148	&	1\\
56414.606088	&	-30.89055	&	0.00143	&	1\\
56415.537112	&	-30.88813	&	0.00247	&	1\\
56415.629409	&	-30.89121	&	0.00169	&	1\\
56416.659510	&	-30.89153	&	0.00130	&	1\\
56451.481081	&	-30.88581	&	0.00104	&	1\\
56452.498147	&	-30.88507	&	0.00110	&	1\\
56454.504932	&	-30.88830	&	0.00123	&	1\\
56455.500043	&	-30.88895	&	0.00149	&	1\\
56458.523950	&	-30.88758	&	0.00105	&	1\\
56673.866608	&	-30.89142	&	0.00143	&	1\\
56691.786236	&	-30.88944	&	0.00116	&	1\\
56692.804280	&	-30.89116	&	0.00126	&	1\\
56693.813456	&	-30.88891	&	0.00110	&	1\\
56694.854936	&	-30.88904	&	0.00101	&	1\\
56695.814952	&	-30.88857	&	0.00119	&	1\\
56696.774954	&	-30.88998	&	0.00116	&	1\\
56697.788917	&	-30.88485	&	0.00204	&	1\\
56712.791363	&	-30.89005	&	0.00116	&	1\\
56713.781551	&	-30.88995	&	0.00124	&	1\\
56714.770625	&	-30.88473	&	0.00263	&	1\\
56715.784792	&	-30.88797	&	0.00118	&	1\\
56716.764372	&	-30.88684	&	0.00130	&	1\\
56717.787019	&	-30.88737	&	0.00162	&	1\\
56718.737567	&	-30.88565	&	0.00136	&	1\\
56719.746495	&	-30.88772	&	0.00110	&	1\\
56720.776348	&	-30.88973	&	0.00130	&	1\\
56721.840783	&	-30.88778	&	0.00129	&	1\\
56722.767771	&	-30.88847	&	0.00119	&	1\\
56723.831797	&	-30.88682	&	0.00125	&	1\\
56724.664996	&	-30.88941	&	0.00128	&	1\\
56724.819142	&	-30.88879	&	0.00118	&	1\\
56725.819715	&	-30.88813	&	0.00123	&	1\\
56726.705643	&	-30.88647	&	0.00126	&	1\\
56726.808723	&	-30.88786	&	0.00108	&	1\\
56727.745772	&	-30.88608	&	0.00115	&	1\\
56728.721606	&	-30.88951	&	0.00113	&	1\\
56729.726790	&	-30.89097	&	0.00110	&	1\\
56740.679336	&	-30.90639	&	0.03978	&	1\\
56740.745134	&	-30.89164	&	0.00107	&	1\\
56741.657045	&	-30.89344	&	0.00108	&	1\\
56742.695366	&	-30.89306	&	0.00123	&	1\\
56745.614222	&	-30.88996	&	0.00121	&	1\\
56746.716007	&	-30.88856	&	0.00117	&	1\\
56763.638265	&	-30.88783	&	0.00127	&	1\\
56764.608497	&	-30.89042	&	0.00107	&	1\\
56765.530894	&	-30.88836	&	0.00138	&	1\\
56766.534987	&	-30.88642	&	0.00131	&	1\\
56767.619700	&	-30.88468	&	0.00139	&	1\\
56768.627806	&	-30.88677	&	0.00101	&	1\\
56778.540910	&	-30.88720	&	0.00103	&	1\\
56779.600943	&	-30.88803	&	0.00114	&	1\\
56781.530809	&	-30.88966	&	0.00124	&	1\\
56782.511550	&	-30.89000	&	0.00118	&	1\\
56783.631955	&	-30.88936	&	0.00146	&	1\\
56784.587602	&	-30.88998	&	0.00116	&	1\\
56822.454912	&	-30.88729	&	0.00203	&	1\\
56823.454293	&	-30.88836	&	0.00165	&	1\\
56826.493163	&	-30.88722	&	0.00155	&	1\\
56827.493656	&	-30.88811	&	0.00122	&	1\\
56828.501242	&	-30.88858	&	0.00108	&	1\\
56837.492926	&	-30.88681	&	0.00144	&	1\\
56838.474070	&	-30.88716	&	0.00111	&	1\\
56839.478407	&	-30.88815	&	0.00137	&	1\\
56857.460518	&	-30.88438	&	0.00142	&	1\\
56858.477550	&	-30.88551	&	0.00119	&	1\\
57018.844391	&	-30.89066	&	0.00125	&	1\\
57019.868509	&	-30.89089	&	0.00119	&	1\\
57020.843780	&	-30.89092	&	0.00117	&	1\\
57021.844886	&	-30.88909	&	0.00130	&	1\\
57022.874432	&	-30.88835	&	0.00121	&	1\\
57044.845249	&	-30.88874	&	0.00129	&	1\\
57045.860911	&	-30.88990	&	0.00088	&	1\\
57046.828490	&	-30.89055	&	0.00104	&	1\\
57047.803082	&	-30.89231	&	0.00109	&	1\\
57048.829182	&	-30.89254	&	0.00133	&	1\\
57049.837861	&	-30.89265	&	0.00129	&	1\\
57050.823874	&	-30.89159	&	0.00105	&	1\\
57051.845202	&	-30.88941	&	0.00101	&	1\\
57052.837984	&	-30.88863	&	0.00115	&	1\\
57053.822872	&	-30.88650	&	0.00131	&	1\\
57055.827529	&	-30.89145	&	0.00134	&	1\\
57056.823768	&	-30.89151	&	0.00103	&	1\\
57057.817621	&	-30.88991	&	0.00136	&	1\\
57063.818028	&	-30.88817	&	0.00122	&	1\\
57064.868448	&	-30.88906	&	0.00107	&	1\\
57065.836178	&	-30.89025	&	0.00115	&	1\\
57066.729781	&	-30.89169	&	0.00167	&	1\\
57075.795269	&	-30.88753	&	0.00108	&	1\\
57076.810866	&	-30.88904	&	0.00097	&	1\\
57077.795256	&	-30.88760	&	0.00143	&	1\\
57078.810387	&	-30.88767	&	0.00108	&	1\\
57079.752863	&	-30.88821	&	0.00127	&	1\\
57080.788778	&	-30.88635	&	0.00158	&	1\\
57082.794062	&	-30.88618	&	0.00119	&	1\\
57085.791436	&	-30.88668	&	0.00102	&	1\\
57100.761528	&	-30.89111	&	0.00131	&	1\\
57101.701955	&	-30.88956	&	0.00189	&	1\\
57102.669638	&	-30.88868	&	0.00200	&	1\\
57117.697316	&	-30.89240	&	0.00132	&	1\\
57135.656832	&	-30.88756	&	0.00091	&	1\\
57137.559269	&	-30.88902	&	0.00112	&	1\\
57138.658439	&	-30.88755	&	0.00122	&	1\\
57139.662591	&	-30.88568	&	0.00123	&	1\\
57146.687254	&	-30.88700	&	0.00148	&	1\\
57211.512772	&	-30.89313	&	0.00137	&	2\\
57405.772480	&	-30.89452	&	0.00087	&	2\\
57413.766190	&	-30.89358	&	0.00113	&	2\\
57423.748994	&	-30.89608	&	0.00097	&	2\\
57424.726953	&	-30.89226	&	0.00091	&	2\\
57425.788480	&	-30.89341	&	0.00096	&	2\\
57446.740198	&	-30.88723	&	0.00081	&	2\\
57447.698608	&	-30.88601	&	0.00104	&	2\\
57448.660142	&	-30.88604	&	0.00119	&	2\\
57450.646377	&	-30.88654	&	0.00116	&	2\\
57451.638971	&	-30.88771	&	0.00085	&	2\\
57456.745843	&	-30.88835	&	0.00119	&	2\\
57470.797926	&	-30.88901	&	0.00138	&	2\\
57472.802257	&	-30.89190	&	0.00095	&	2\\
57479.768521	&	-30.89012	&	0.00097	&	2\\
57486.592394	&	-30.88839	&	0.00119	&	2\\
57486.751671	&	-30.88854	&	0.00116	&	2\\
57487.557776	&	-30.88713	&	0.00115	&	2\\
57487.714020	&	-30.88698	&	0.00187	&	2\\
57488.609334	&	-30.88882	&	0.00086	&	2\\
57488.773309	&	-30.88808	&	0.00106	&	2\\
57489.577410	&	-30.88887	&	0.00157	&	2\\
57504.718573	&	-30.88342	&	0.00509	&	2\\
\end{longtable}}
\twocolumn

\end{document}